\documentclass[aps, pra,superscriptaddress,twocolumn,a4paper]{revtex4-1}
\usepackage{bm}
\usepackage{appendix}
\usepackage{graphicx}
\usepackage{amsmath}
\usepackage{amssymb}%
\usepackage{color}
\usepackage{lipsum}
\usepackage[colorlinks=true,citecolor=blue,urlcolor=blue]{hyperref}
\begin{document}

\title{Enhanced Fulde-Ferrell-Larkin-Ovchinnikov and Sarma superfluid states near an orbital Feshbach resonance}
\author{Dongyang Yu}
\affiliation{Beijing National Laboratory for Condensed Matter Physics, Institute of
Physics, Chinese Academy of Sciences, Beijing 100190, China}
\affiliation{School of Physical Sciences, University of Chinese Academy of Sciences,
Beijing 100190, China}
\author{Wei Zhang}
\email{wzhangl@ruc.edu.cn}
\affiliation{Department of Physics, Renmin University of China, Beijing 100872, China}
\affiliation{Beijing Key Laboratory of Opto-electronic Functional Materials and Micro-nano Devices,
Renmin University of China, Beijing 100872, China}
\author{Wu-Ming Liu}
\email{wmliu@iphy.ac.cn}
\affiliation{Beijing National Laboratory for Condensed Matter Physics, Institute of
Physics, Chinese Academy of Sciences, Beijing 100190, China}
\affiliation{School of Physical Sciences, University of Chinese Academy of Sciences,
Beijing 100190, China}
\affiliation{Songshan Lake Materials Laboratory, Dongguan, Guangdong 523808, China}

\date{\today}

\begin{abstract}
We investigate the Fulde-Ferrell-Larkin-Ovchinnikov (FFLO) and Sarma superfluid states in alkaline-earth-like $^{173}$Yb atomic gases near an orbital Feshbach resonance at zero temperature with population imbalances in both the open and closed channels.
We find that in uniform space both the Fulde-Ferrell and Sarma states are greatly enhanced by the spin-exchange interaction in the Bardeen-Cooper-Schrieffer side of the Feshbach resonance. While trapped in a harmonic potential, a cloud of long-lived $^{173}$Yb atomic gas with small fraction of electronically-excited state population can stabilize not only the Sarma states with both the one and two Fermi surfaces but also the Fulde-Ferrell state, and leave detectable structures in the distribution of polarizations of different bands.
As the degenerate $^{173}$Yb cloud is readily available and the signatures predicted can be easily detected in \textit{in-situ} and/or time-of-flight images, our findings are helpful to realize and detect the long-sought FFLO and Sarma states in experiments.
\end{abstract}
\maketitle

\section{introduction}
The pairing between fermions and the resulting superfluid (SF) phases in the presence of a Zeeman energy (ZE) is one of the key questions in multidisciplinary fields of physics. Two well-known prototypes of unconventional SF, the inhomogeneous Fulde-Ferrell-Larkin-Ovchinnikov (FFLO) phase~\cite{FF1964,LO1964,casalbuoni2004} and the homogenous Sarma (also referred as breached-pair) state~\cite{vwliu2003,forbes2005}, have been proposed and extensively studied in condensed matter physics~\cite{barzykin2007,barzykin2009,subasi2010,cho2017}, cold atomic gas~\cite{carlson2005,zwierlein2006,shin2006,sheehy2006,wyi2006-1,wyi2006-3,parish2007-1,cherng2007,orso2007,yliao2010,radzihovsky2010,strack2014,boettcher2015-1,
revelle2016}
and chromodynamics~\cite{gubankova2003,casalbuoni2004}.
As the ZE induces mismatch between Fermi surfaces, pairing can either take place between fermions near shifted Fermi surfaces to form FFLO states with Cooper pairs of non-zero center-of-mass momentum, or involve particles inside the larger Fermi sea to establish Sarma phase with gapless excitations.

For alkaline-metal atomic gases near a magnetic Feshbach resonance (MFR), the FFLO phase is predicted to be stable only within a narrow sliver of parameter space in the Bardeen-Cooper-Schrieffer (BCS) regime~\cite{zwierlein2006,sheehy2006,radzihovsky2010,revelle2016}, while the Sarma phase with one Fermi surface (FS) exists only in the Bose-Einstein-condensate (BEC) regime and the one with two FSs is always unstable~\cite{carlson2005,sheehy2006,wyi2006-1,wyi2006-3,parish2007-1,radzihovsky2010}. Although earlier works suggest that either FFLO or Sarma SF can be realized under assistance of various mechanisms, e.g., spin-orbit coupling~\cite{yjlin2011,fwu2013,liao2012}, multiband effect of optical lattices~\cite{ksun2013,he2009,sywang2017,huhtinen2018,tylutki2018}, and low-dimensionality~\cite{orso2007,yliao2010}, a crystal sharp evidence for the realization of FFLO and Sarma states is still hindered by experimental difficulties and technique limitations. Recently, a new kind of Feshbach resonance referred as orbital Feshbach resonance (OrbFR) is theoretically proposed and experimentally verified in alkaline-earth-like atomic gases~\cite{rzhang2015,hofer2015,pagano2015}. Comparing to MFR, an OrbFR system involves four atomic levels~\cite{cherng2007,boettcher2015-2} and hence two independent ZEs in the open and closed channels, as well as a spin-flip inter-channel interaction. These characteristics bring new aspects to many physical properties, including the SF transition temperature~\cite{xu2016}, collective excitations~\cite{he2016,yczhang2017,zou2018}, polaron-molecule transition~\cite{jgchen2016}, and topological states with spin-orbit coupling~\cite{mancini2015,livi2016,swang2017,xfzhou2017,iemini2017}. Specifically, Ref.~\cite{zou2018} proposes an emergent Sarma state in large but equal ZEs ($h_o=h_c$) in the open and closed channels with an artificial choice of parameters, while Ref.~\cite{iskin2017} studies the pair-breaking effect of finite temperature in a harmonic trap without ZE.

Here we investigate the pairing states in degenerate $^{173}$Yb gases as the unique OrbFR system up to date, at zero temperature with tunable ZEs using realistic experimental settings and physical parameters.
The atoms are prepared in the lowest two electronic manifolds $|^1S_0\rangle$ (denoted by orbital $|g\rangle$) and $|^3P_0\rangle$ ($|e\rangle$) with nuclear spins $m_\uparrow$ (labeled by pseudo-spin $|\uparrow\rangle$) and $m_\downarrow$ ($|\downarrow\rangle$). The open channel is composed of $|o,\uparrow\rangle \equiv |e,\uparrow\rangle$ and $|o,\downarrow\rangle\equiv|g,\downarrow\rangle$, while the closed channel involves $|c,\uparrow\rangle \equiv |g,\uparrow\rangle$ and $|c,\downarrow\rangle\equiv|e,\downarrow\rangle$.
The pairing states in the open and closed channels can in principle be different. For a 3-dimensional (3D) uniform system, we find by employing a mean-field approach that among all possible combinations of SF states in different channels, only four of them are stabilized as listed in Fig.~\ref{Fig1}. Specifically, at least one of the two channels (labeled by the subscript $o$ or $c$) must be in the BCS phase regardless of the choice of ZEs. In the limiting case of $h_c=0$, both FFLO and Sarma states are stabilized in a rather large regime of BCS side.
We then discuss the effect of an external harmonic trap under the local-density approximation, and show that both the FFLO and Sarma states leave sizable signatures of density distributions~\cite{wyi2006-3,qzhou2009} which can be easily detected by \textit{in-situ} and/or time-of-flight imaging techniques~\cite{oppong2019}.

\section{Model}\label{section-model}
For an OrbFR system, the Hamiltonian in the grand canonical ensemble takes the form,
\begin{eqnarray}\label{H}
\hat H&=&\sum_{\bm{k}} \sum_j^{o,c}\sum_\sigma^{\uparrow,\downarrow} \xi_{k,j,\sigma}\hat c^\dagger_{\bm{k},j,\sigma}{\hat c_{\bm{k},j,\sigma}}\nonumber\\
&+&\sum^{o,c}_{i,j}U_{ij}\sum_{\bm{q},\bm{k},\bm{k}^\prime}{\hat c}^\dagger_{\bm{q}-\bm{k},i,\downarrow}{\hat c}^\dagger_{\bm{q}+\bm{k},i,\uparrow}{\hat c}_{\bm{q}+\bm{k}^\prime,j,\uparrow}{\hat c}_{\bm{q}-\bm{k}^\prime,j,\downarrow},
\end{eqnarray}
where $\xi_{\bm{k},j,\sigma}=\epsilon_{\bm{k}} -\mu_j+\xi_\sigma h_j$, the kinetic energy $\epsilon_{\bm{k}}=\hbar^2|{\bm k}|^2/2M$ with $M$ the atomic mass, the chemical potentials $\mu_o=\mu$ and $\mu_c=\mu-\delta/2$. The inter-channel detuning $\delta=\delta_g \mu_N \Delta_m |\bm{B}|$ is well controlled by an external magnetic field $\bm B$ with $\delta g$ the nonzero differential Land{\'e} $g$ factor between $|g\rangle$ and $|e\rangle$, $\mu_N$ the nuclear Bohr magneton and $\Delta_m=m_\uparrow-m_\downarrow$. $\hat c_{{\bm k},j,\sigma}$ ($\hat c^\dagger_{{\bm k},j,\sigma}$) is the annihilation (creation) operator of atom with momentum $\bm{k}$ and spin $\sigma$ in the channel $j$, and an effective ZE $h_j$ ($\xi_\sigma=\pm 1$ for spin-up and spin-down) is additionally applied by introducing intra-channel population imbalances that are one-to-one mapping to the experimentally tunable imbalances in the degree of nuclear spin and electronic orbit . The intra-channel coupling strength is symmetric $U_{oo}=U_{cc}=U_0$ and the inter-channel spin-exchange interaction $U_{oc}=U_{co}=U_1$ with $U_{0}=(U_+ + U_-)/2$ and $U_{1}=(U_+ - U_-)/2$. Here, $U_{\pm}$ are related to the corresponding $s$-wave scattering length $a_{s\pm}$ via the conventional renormalization relation~\cite{renormalization}.
By increasing $\delta$ from zero (set $\delta\geq0$), the effective scattering length $a_s$ between particles in the open channel is tuned from the BEC regime to the resonance point  at $\delta_{\rm res}=4\hbar^2/M(a_{s+}+a_{s-})^2$, and to the deep BCS regime~\cite{rzhang2015}. Notice that it is intrinsic that the deep BEC regime can not be reached in OrbFR.

For simplicity, we consider here only the Fulde-Ferrell (FF) state where all Cooper pairs acquire the same momentum. Although the
Larkin-Ovchinnikov state with two components of opposite momenta is in principle more stable than the FF state, they are
qualitatively consistent around the phase boundaries~\cite{radzihovsky2010,kinnunen2018}.
Due to the momentum conservation, the momentum of Cooper pairs $2\hbar\bm{Q}$ are the same in both channels, and the order parameters
in different channels can be written as
\begin{eqnarray}\label{order-para}
\Delta_{o}=\sum_{\bm k}( U_{0} \langle {\hat c}_{\bm{Q}+\bm{k},o,\uparrow}{\hat c}_{\bm{Q}-\bm{k},o,\downarrow} \rangle
+U_{1} \langle {\hat c}_{\bm{Q}+\bm{k},c,\uparrow} {\hat c}_{\bm{Q}-\bm{k},c,\downarrow} \rangle),\nonumber\\
\Delta_{c}=-\sum_{\bm k}( U_{1} \langle {\hat c}_{\bm{Q}+\bm{k},o,\uparrow}{\hat c}_{\bm{Q}-\bm{k},o,\downarrow} \rangle
+U_{0} \langle {\hat c}_{\bm{Q}+\bm{k},c,\uparrow} {\hat c}_{\bm{Q}-\bm{k},c,\downarrow} \rangle),\nonumber\\
\end{eqnarray}
Within the mean-field treatment, the order parameters $\Delta_{o/c}$ and the wavevector of Cooper pairs $|{\bm Q}|$ are self-consistently determined by the saddle-point equations which extremize the thermodynamical potential $\delta G$ of Hamiltonian Eq.~(\ref{H}) with respect to the normal phase~\cite{radzihovsky2010} (Appendix~\ref{append-a}).
As $\delta G$ is an even function for both ZEs, we assume $h_{o/c}$ are positive definite without loss of generality.

The saddle-point equations have multiple solutions for the OrbFR system. One is an in-phase solution where the order parameters $\Delta_o$ and $\Delta_c$ have the same phase, while the other one is an out-of-phase solution with a $\pi$ phase difference. For $^{173}$Yb, the two scattering lengths are positive and satisfy $a_{s+}\gg a_{s-}>0$ (see below). Thus, the absolute ground state is the in-phase solution which corresponds to a rather trivial SF phase of deeply bound dimers. The metastable out-of-phase solution can be tuned through the BEC-BCS crossover~\cite{iskin2016,cappellini2018} and is proved to be both mechanically and dynamically stable in the case without population imbalance~\cite{he2016}. We focus on this out-of-phase metastable solution, assuming it is also stable in the presence of population imbalance, and consider it as the `{\it ground}' state (Appendix~\ref{append-a}) unless specified. We also assume without loss of generality that $\Delta_o \ge 0$ and $\Delta_c \le 0$.

\begin{figure}[tbp]
  \centering
  \includegraphics[width=8.5cm]{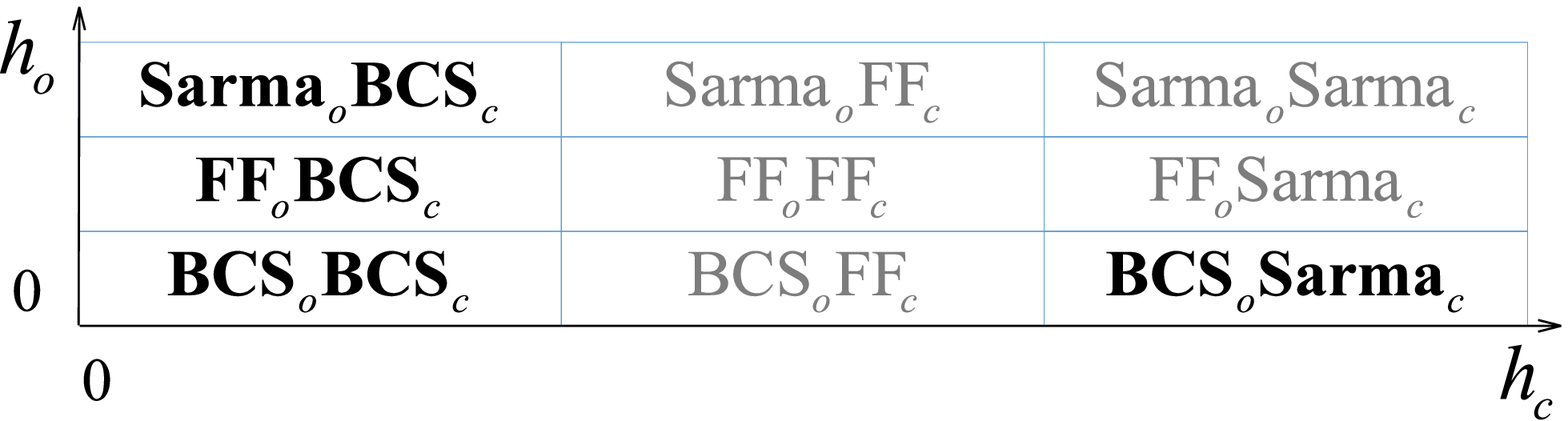}\\
    \caption{(Color online) Summary of all possible SF phases in the presence of finite ZEs in the open and closed channels. The combined symbol ${\rm X}_{o}{\rm Y}_{c}$ represents that the open (closed) channel is X (Y) phase with X or Y$\equiv$ BCS, FF and Sarma. The combinations labeled in black bold (grey normal) are (not) observed in systems with realistic parameters.}
    \label{Fig1}
\end{figure}

In the presence of ZEs, the pairing state in each channel can be chosen from the BCS, FF, and Sarma states, which are characterized by $|\bm{Q}| = 0$ with no polarization, $|\bm{Q}| \neq 0$, and $|\bm{Q}| = 0$ with polarization, respectively. There are hence nine different combinations as depicted in Fig.~\ref{Fig1}. However, in a fairly large parameter regime which is realistic in experiments, five of these possibilities (gray in Fig.~\ref{Fig1}) are ruled out from the `{\it ground}' state as defined above. In particular, the pairing state in the closed channel is most likely to be in the BCS phase, except in the limiting case of large $h_c$ and small $h_o$ where the Sarma phase (only one FS in most regime except around $\delta=0$) is stabilized in the closed channel while the open channel is a BCS state (i.e., BCS$_o$Sarma$_c$). This result can be understood by noticing that as the closed channel is highly detuned, the low-lying states in the open channel can assist pairing in the closed channel via the spin-exchange interaction, in return favor the conventional BCS state which has no modulation or nodal structure. Another consequence of the inter-channel interaction is that the two order parameters $\Delta_{o/c}$ approach zero simultaneously with increasing ZEs, showing that the superfluidity in both channels are intimately connected with each other as suggested in a finite-temperature analysis of the population balanced case~\cite{xu2016}.

The pairing state in the open channel, however, possesses a rich phase diagram with all three possibilities of BCS, FF, and Sarma states when $h_c$ is weak. Thus, next we focus on the limiting case of $h_c = 0$ and investigate the phase diagram by varying $h_o$. The general scenario in the presence of both ZEs will be discussed in the section~\ref{section-ZEs}. We remark here that the ZEs $h_{o/c}$ in both channels can be independently tuned by the population imbalances in both channels ( $P_{j}=(N_{j,\downarrow}-N_{j,\uparrow})/(N_{j,\downarrow}+N_{j,\uparrow})$ for $j\equiv {o,c}$ ), or equivalently by the imbalances in the two electronic orbits $|g\rangle$ and $|e\rangle$,
\begin{align}
P_g=\frac{N_{g,\downarrow}-N_{g,\uparrow}}{N_{g,\downarrow}+N_{g,\uparrow}}=\frac{(1+P_o)(1-F_c)-F_c(1-P_c)}{(1+P_o)(1-F_c)+F_c(1-P_c)},\\
P_e=\frac{N_{e,\downarrow}-N_{e,\uparrow}}{N_{e,\downarrow}+N_{e,\uparrow}}=\frac{(1+P_c)F_c-(1-P_o)(1-F_c)}{(1+P_c)F_c+(1-P_o)(1-F_c)}.
\end{align}
The fraction of the $|e\rangle$ state $F_e$ is related to the fraction of the closed channel $F_c=(N_{c,\downarrow}+N_{c,\uparrow})/N$ by
\begin{equation}
F_e=\frac{N_{e,\downarrow}+N_{e,\uparrow}}{N}=\frac{ (1-P_o)(1-F_c)+(1+P_c)F_c }{2},
\end{equation}
both of them are self-consistently determined by the saddle-point equations, here N is the total number of atoms.

\section{BEC-BCS crossover at $h_c=0$}\label{section-crossover}
In the absence of population imbalance, particles in the closed channel are either in the BCS$_c$ phase with finite order parameters $\Delta_{o/c}$, or in the unpolarized normal (vacuum) state with $\Delta_{o/c}=0$ for $\delta/2 < \mu$ ($\delta/2 \ge \mu$). Thus, in the following discussion we focus only on the open channel without mentioning the state of closed channel unless necessary. To get a clear connection with the case without population imbalance, we use a dimensionless quantity ${\hat h}_o \equiv h_o/\Delta_o(h_o=0, h_c = 0)$ to measure the ZE of open channel through rescaling $h_o$ by the corresponding order parameter $\Delta_o$ when $h_o = h_c = 0$ for given inter-channel detuning $\delta$ and chemical potential $\mu$. As a concrete example, we take $a_{s+} = 1900 a_0$ and $a_{s-} = 210 a_0$~\cite{scazza2014,hofer2015,pagano2015} for $^{173}$Yb with $a_0$ the Bohr radius, and consider a gas in 3D uniform space with mean total number density $n_0=5.2\times 10^{15} \rm cm^{-3}$, which corresponds to $\left(k_{\rm F} a_{s+}\right)^{-1}=0.062$ and $\left(k_{\rm F} a_{s-}\right)^{-1}=0.588$ with the Fermi wavevector $k_{\rm F}={(3\pi^2 n_0)}^{1/3}$.
\begin{figure}
  \centering
  \begin{minipage}[b]{0.5\textwidth}
  \centering
  \includegraphics[width=8.5cm]{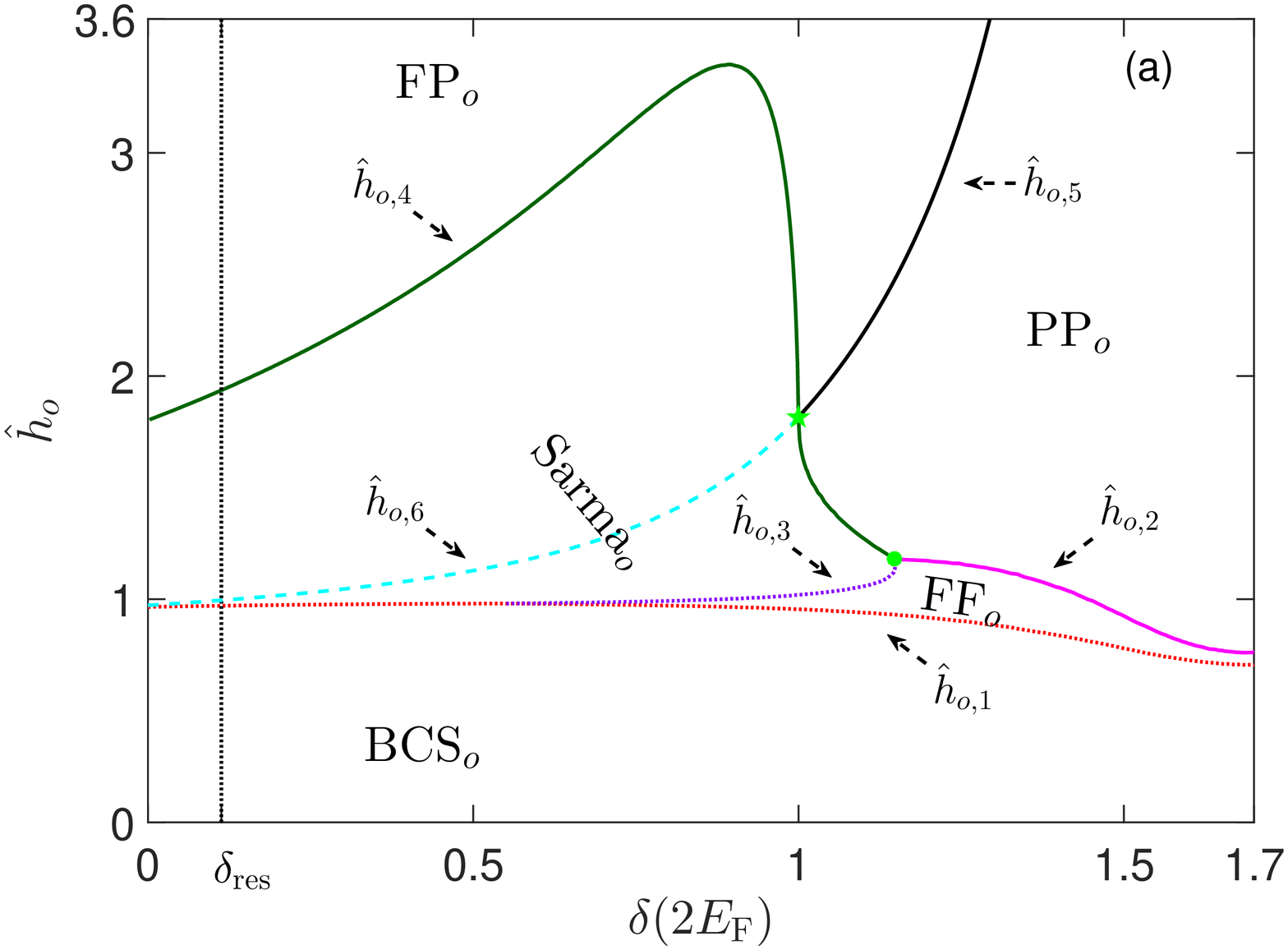}
  \end{minipage}
  \par\vspace{0pt}
  \begin{minipage}[b]{0.5\textwidth}
  \centering
  \includegraphics[width=8.5cm]{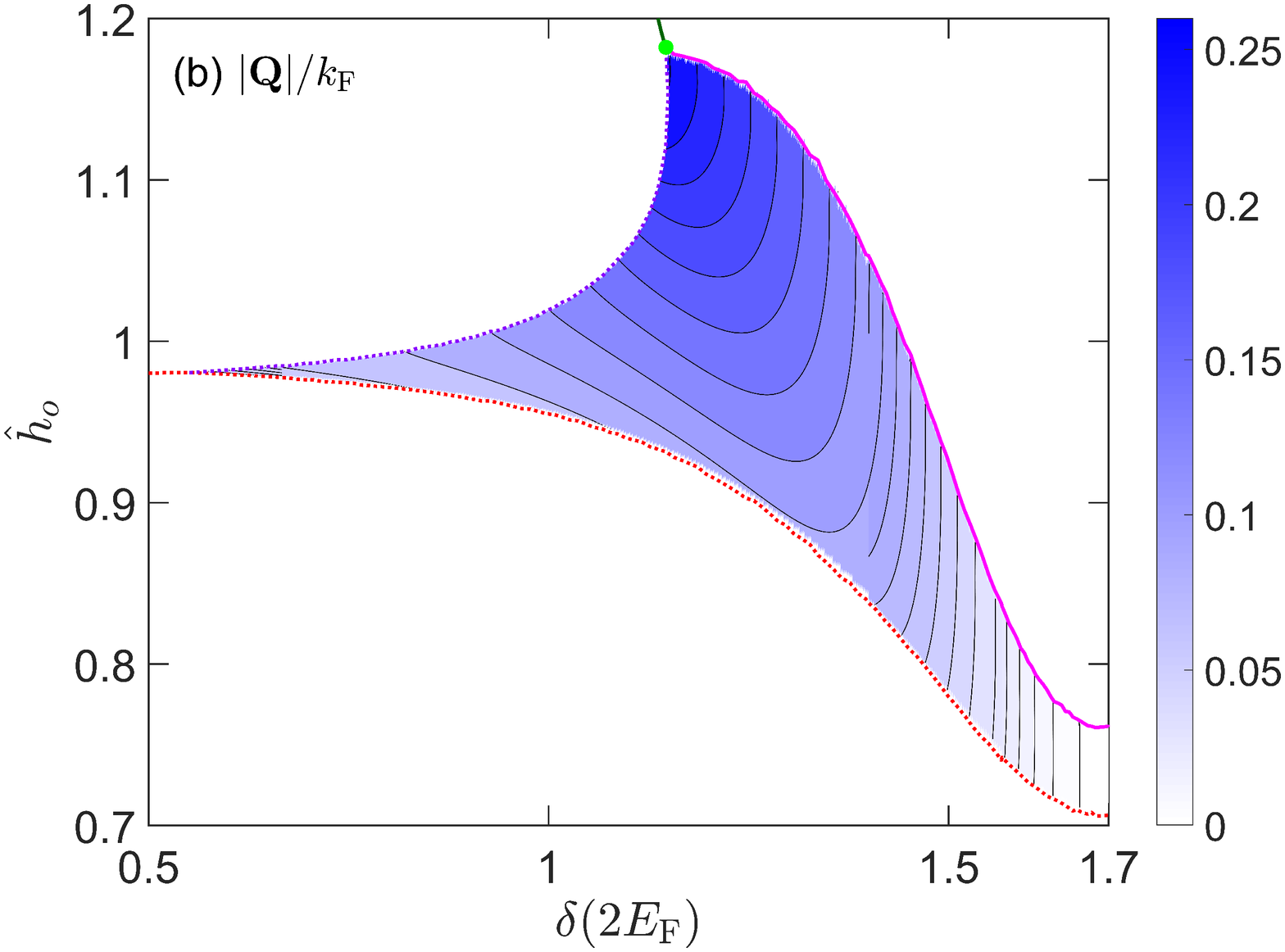}
  \end{minipage}
  \par\vspace{0pt}
  \caption{(Color online) (a) Zero temperature phase diagram in the $\delta$--$\hat{h}_o$ ($\hat{h}_o$ is dimensionless) plane with $h_c=0$ and $\mu = E_{\rm F}$ (in 3D uniform space). Three distinctive SF phases of the open channel--BCS$_o$, Sarma$_o$ and FF$_o$--are present while the closed channel remains in the BCS$_c$ phase. PP$_o$ (FP$_o$) stands for the partially (fully)-polarized normal state. The thick dotted/solid lines (labeled as ${\hat h}_{o,i}$ for $i=1,\cdots, 5$) represent the first/second order quantum phase transition (Appendix~\ref{append-b}), which are hinged by tricritical (green circle) and tetracritical (green pentagram) points. The thin black dotted line indicates the resonance point $\delta_{\rm res}$ while the thin cyan dashed line ${\hat h}_{o,6}$ stands for the topological transition between Sarma$_o$ states with one and two FSs. (b) Contour plot of the wavevector $|{\bm Q}|$ in the FF$_o$ regime.}\label{Fig2}
\end{figure}

In Fig.~\ref{Fig2}(a), we depict the zero temperature phase diagram by varying $\delta$ and ${\hat h}_o$ for a fixed $\mu = E_{\rm F}$. Generally speaking, the system is in the BCS$_o$ state when ${\hat h}_o$ is small, becomes Sarma$_o$ or FF$_o$ for moderate ${\hat h}_o$ depending on the detuning, and eventually turns into a fully (FP$_o$) or partially (PP$_o$) polarized normal state for even larger ${\hat h}_o$. In the BEC regime of $\delta < \delta_{\rm res}$, the BCS gap $\Delta_o$ becomes larger than the chemical potential $\mu$ due to the formation of tightly bound molecules~\cite{cappellini2018}. As a result, the critical ZE ${\hat h}_{o,1} \approx 1$, beyond which the Sarma$_o$ phase can be stabilized by pairing few atoms around zero kinetic energy (also paring around the larger FS of the open channel). Actually the Sarma$_o$ phase possesses two FSs within the parameter region ${\hat h}_o < {\hat h}_{o,6}$, and features one FS for ${\hat h}_{o} > {\hat h}_{o,6}$ (Appendix~\ref{append-c}). Note that the line of ${\hat h}_{o,6}$ remains above ${\hat h}_{o,1}$ throughout the OrbFR , with a small interval 0.07 at $\delta = 0$. In the BCS regime of $\delta > \delta_{\rm res}$, the critical ZE ${\hat h}_{o,1}$ decreases from the limiting value of unity, and the Sarma$_o$ state is gradually replaced by the FF$_o$ state with nonzero wavevector $|\bm Q|$. In particular, in the BCS limit with $\delta/(2E_{\rm F})\gtrsim 1.7$, the lower (${\hat h}_{o,1}$) and upper (${\hat h}_{o,2}$) boundaries of FF$_o$ saturate to $1/\sqrt{2} \approx 0.756$, in consistency with the results in conventional MFR~\cite{he2006,sheehy2006,zou2018}.

A distinct feature of the phase diagram Fig.~\ref{Fig2}(a) is that both the stable regions of Sarma$_o$ and FF$_o$ are greatly enlarged by evolving from the BEC to the unitarity and to the BCS regime for $\delta \lesssim 2\mu$. This is in stark contrast to the case of broad MFR, where the Sarma and FF phases are both hindered by crossing over from the BEC to the BCS side~\cite{zwierlein2006,sheehy2006,radzihovsky2010}. To understand this result, we notice that when $\delta < 2\mu$, both the open and closed channels are populated, such that the Hamiltonian Eq.~(\ref{H}) can be considered as a two-band model with effectively asymmetric interactions resulting from the finite detuning $\delta$. Thus, as one enhances the asymmetry between the two bands with increasing $\delta$, the Sarma and FF states can be further stabilized by the extra closed channel and the corresponding inter-channel pair tunneling~\cite{he2009,subasi2010,ksun2013,revelle2016,iskin2016,zou2018} induced by spin-exchange interaction (Eq.~\ref{PairTunneling}). On the other hand, when $\delta \gtrsim 2\mu$, the highly detuned closed channel is essentially frozen and the inter-channel pair tunneling is significantly reduced.
As a result, the system gradually reduces to an effective single-band model as for a broad MFR~\cite{lyhe2015}, where the Sarma (FFLO) state becomes unstable (fragile).

The crossover from a two-band to a single-band model can also be observed from the variation of wavevector $|\bm Q|$ of the FF$_o$ state, as shown in Fig.~\ref{Fig2}(b). In the two-band side with $\delta \lesssim 2\mu$, FF$_o$ emerges with a small $|\bm Q| \lesssim 0.05k_{\rm F}$. The increase of $\delta$ within this regime can enhance the inter-channel pair tunneling, such that the system can tolerate a much larger wavevector. After reaching the maximal value of $|\bm Q| \approx0.26k_{\rm F}$ at $\delta/2E_{\rm F} \approx 1.147$, the wavevector starts to drop with $\delta$, and approaches to zero as one would naturally expect for a single-band model in the deep BCS limit~\cite{radzihovsky2010}.

\section{Typical Phase Diagram with both Zeeman energies}\label{section-ZEs}
\begin{figure}
  \centering
  \includegraphics[width=8.5cm]{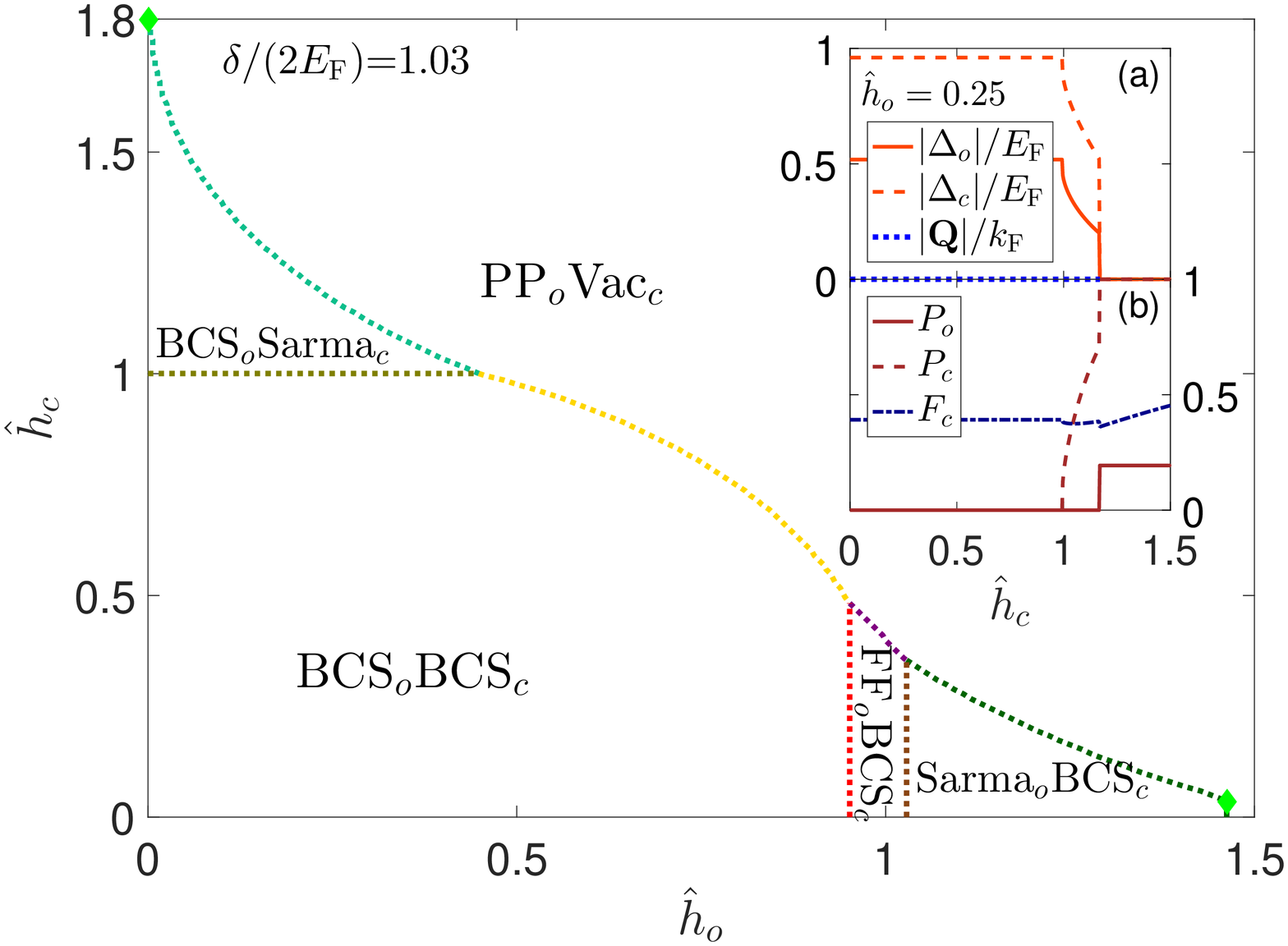}\\
  \caption{(Color online) A typical phase diagram in the dimensionless scaled $\hat{h}_o$--$\hat{h}_c$ plane at zero temperature when $\delta/(2E_{\rm F})=1.03$. Other settings can be found in Fig.~\ref{Fig2} and the symbols are explained in the main text. The inset shows the evolutions along $h_c$ line with fixed ${\hat h}_o=0.25$.} \label{Fig3}
\end{figure}

In the case of both finite ZEs in both channels, $h_o>0$ and $h_c>0$, a typical phase diagram is shown in the dimensionless scaled ZEs $\hat{h}_{o/c}\equiv h_{o/c}/|\Delta_{o/c}(h_o=0,h_c=0)|$ in Fig.~\ref{Fig3} with $\delta/(2 E_{\rm F})=1.03$ and $\mu=E_{\rm F}$.
As one can see, by varying ${\hat h}_o$ and ${\hat h}_c$, all four stable SF phases, including BCS$_o$BCS$_c$, FF$_o$BCS$_c$, Sarma$_o$BCS$_c$ and BCS$_o$Sarma$_c$, are present in the lower-left side of the phase diagram, while in the upper-right corner with quite large ZEs, particles in the open channel is in PP$_o$ state due to ${\hat h}_o<{\hat h}_{o,5}$ and the closed channel is in vacuum (Vac$_c$) because of $\mu_c<0$. This phase diagram suggests a general scenario that the presence of ZE in one channel would reduce the SF order parameter of the trivial BCS phase and consequently suppress pairing in the other channel through inter-channel pair tunneling. We note here that the critical scaled ZE from BCS$_o$BCS$_c$ to BCS$_o$Sarma$_c$ is very close to unity because the effective gap of the closed channel $\Delta_{{\rm eff},c}=\sqrt{\Delta_c^2+\mu_c^2\Theta(-\mu_c)} \cong |\Delta_c|$. From the behaviors of relevant physical quantities shown in the insets of Fig.~\ref{Fig3}, we conclude that all phase transitions in Fig.~\ref{Fig3} labeled by dotted lines are of the first order, except around the points (labeled by green diamond symbols) in the upper-left and the lower-right corners with highly asymmetric ZEs where the transitions are of the second order.

\section{detection}
To facilitate the identification of the exotic states discussed above, we next study the distributions of atoms in an external  3D isotropic harmonic trap $V_{\rm ext}({\bm r}) = M \omega^2 r^2/2$ at zero temperature under the local-density approximation~\cite{qzhou2009}.
Comparing with a broad MFR system, a key characteristic of the OrbFR is that with increasing $h_o$, the BCS$_o$ core can be completely suppressed and replaced by the Sarma$_o$ or FF$_o$ state, therefore leaving detectable features in the distribution of polarizations of relevant bands.
\begin{figure}
  \centering
  \includegraphics[width=8.5cm]{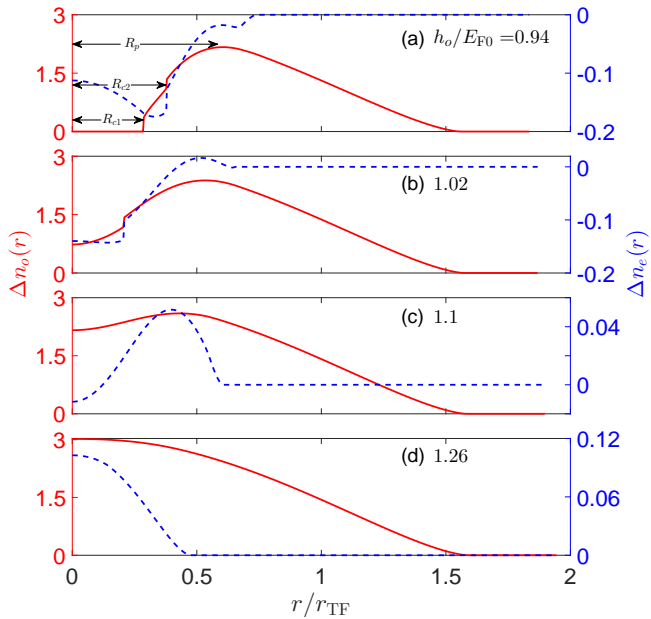}\\
  \caption{(Color online) Four typical polarizations in 3D isotopic harmonic trap, $\Delta n_o(r)$ and $\Delta n_e(r)$, versus the radial radius (the Thomas-Fermi radius $r_{\rm TF}$) in panels (a-d) as $h_o$ varies. The three critical radii of $\Delta n_o(r)$ are demonstrated in panel (a).
  In the plot, we choose a typical $\delta/(2E_{{\rm F}0})=1.03$ in the BCS side with $h_c=0$ and the Fermi energy $E_{{\rm F}0}$ corresponding to total number density $n_0=5.2 \times 10^{15}{\rm cm^{-3}}$.}\label{Fig4}
\end{figure}

\begin{figure}
  \centering
  \includegraphics[width=8.5cm]{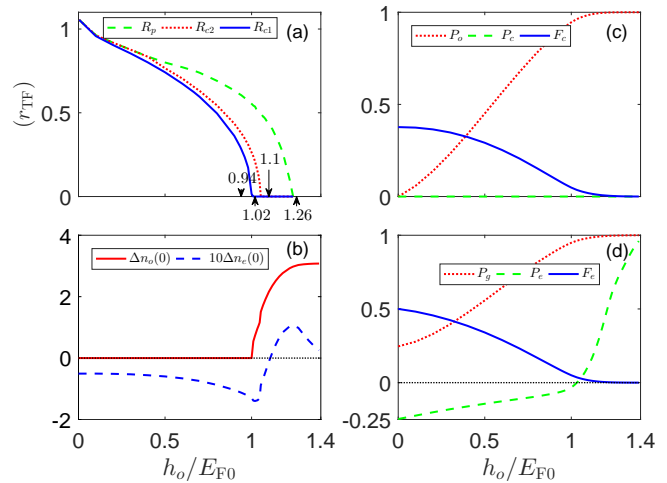}\\
  \caption{(Color online) Evolutions of several measurable quantities are shown as $h_o$ increases, including the three critical radii (a), the polarizations of $\Delta n_{o/e}(r=0)$ in the center of trap (b) and the polarizations and fractions in the degree of channels (c) and electronic orbits (d). All settings are the same as in Fig.~\ref{Fig4}.}\label{Fig5}
\end{figure}

One measurable quantity is the distribution of polarization $\Delta n_{o}(r)=n_{o,\downarrow}(r)-n_{o,\uparrow}(r)$ with $n_{o,\sigma}(r)=\sum_{\bm k} n_{o,\sigma}({\bm k},r)$ along the radial axis, which can be obtained either by counting independently the two density distributions $n_{o,\downarrow}(r)$ and $n_{o,\uparrow}(r)$, or by directly measuring the density difference between $|g\rangle$ and $|e\rangle$ through phase-contrast imaging technique~\cite{shin2006} working at anti-magic wavelength~\cite{hofer2017-thesis} (when $h_c=0$).
Four typical distributions of polarization $\Delta n_{o/e}(r)$~\cite{dne} are displayed in Figs.~\ref{Fig4}(a)-(d).
When $h_o$ is weak, $\Delta n_o(r)$ shows clear features with an empty core (up to a critical radius $R_{c1}$), a secondary jump (at position $R_{c2}$) and a peak (at $R_{p}$), due to the different polarizations of various SF phases (Appendix~\ref{append-b}).
The empty core and the secondary jump, corresponding respectively to the BCS$_o$ and FF$_o$ states, disappear when $h_o$ is large enough. The evolution of the shell structures of a harmonically-trapped gas can be seen from the critical radii shown in Fig.~\ref{Fig5}(a) and the polarizations $\Delta n_o(r=0)$ or $\Delta n_e(r=0)$ at trap center as in Fig.~\ref{Fig5}(b). In both panels, clear phase transitions are demonstrated when $h_o \approx E_{{\rm F}0}$.

We emphasize that as $h_o$ can suppress the superfluidity through inter-channel pair tunneling, the excited fraction $F_e$ and closed channel fraction $F_c$ are both small at large $h_o$ as in Fig.~\ref{Fig5}(c-d). Although the inelastic collisions between two $|e\rangle$ atoms will induce severe atom loss and heating, thereby preventing a high fraction of the excited state $|e\rangle$, a quasi-two-dimensional cloud of highly polarized but long-lived $^{173}$Yb atomic gas is still available experimentally with $F_e$ as large as 22$\%$ at temperature as low as 0.14$E_{\rm F}$~\cite{oppong2019}. Our results hence suggest a practical scheme to detect the FFLO and Sarma signatures in trapped alkaline-earth-like gases.

\begin{figure}
  \centering
  \includegraphics[width=8.5cm]{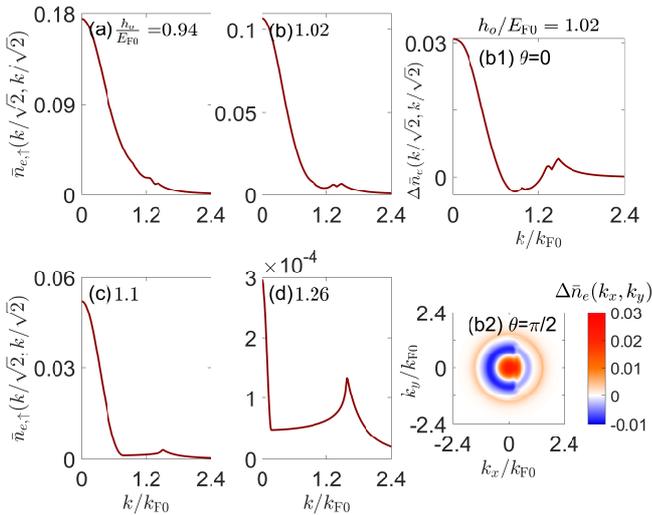}\\
  \caption{(Color online) Columnar-integrated momentum distributions ${\bar n}_{e\uparrow}(k_x,k_y)$ along the radial axis in panels (a-d) when $\theta=0$, and $\Delta {\bar n}_e(k_x,k_y)$ in panels (b1-b2) for $h_o/E_{{\rm F}0}=1.02$. Other parameters are the same as in Fig.~\ref{Fig4}. }\label{Fig6}
\end{figure}

Another accessible quantity is the columnar-integrated density distributions in momentum space for the $|e,\sigma \rangle$ state ${\bar n}_{e,\sigma}(k_x,k_y)=k_{{\rm F}0}^2/(2\pi^2N)\int dk_z \int dr r^2 n_{e,\sigma}({\bm k},r)$ or the polarization $\Delta {\bar n}_e(k_x, k_y)={\bar n}_{e,\downarrow}-{\bar n}_{e,\uparrow}$, which is about angle $\theta$ with the FF wavevector ${\bm Q}$ and can be easily obtained by a procedure similar to the \textit{in-situ} imaging.
For alkaline-metal atoms around a broad MFR, the Sarma state keeps stable only in the BEC side therefore with one FS, whose signature can be detected by spin-selective time-of-flight imaging~\cite{wyi2006-3} and the one of FFLO state is usually smeared out by the robust BCS state in the trap center~\cite{zwierlein2006,revelle2016}.
However, for the OrbFR system in the BCS side, we identify clear signatures not only for Sarma state with one and two FS(s), but also for the anisotropic FF state.
In Figs.~\ref{Fig6}(a-d), we show the momentum density distribution $n_{e,\uparrow}$ along the radial momentum corresponding to scenarios in Figs.~\ref{Fig4}(a-d), as the system acquires an axial symmetry when $\theta=0$.
As we can see, a non-monotonic valley structure with a double- or single-peak gradually emerges around the Fermi wavevector $k_{{\rm F}0}$, stemming from the fully-empty shell (ball) of $|e,\uparrow\rangle$ between the two FSs (below the one FS) in the Sarma$_o$ with two (one) FSs and the depletion of BCS$_o$ or FF$_o$ state, while the robust core around zero momentum eventually disappears because it is dominated by the Sarma$_o$ state with one FS as in Fig.~\ref{Fig6}(d).
The above signatures can be seen more apparently by measuring the polarization $\Delta {\bar n}_e(k_x,k_y)$
in the $|e\rangle$ state (Appendix~\ref{append-d}), exemplified as in Fig.~\ref{Fig6}(b1-b2) with $h_o/E_{{\rm F}0}=1.02$.
Notice that the fully empty shell or ball of the Sarma state enhances the anisotropy induced by the FF$_o$ state, which is maximized when $\theta=\pi/2$ as in Fig.~\ref{Fig6}(b2).

\section{conclusion}
We study exotic pairing phases in ultracold gases of $^{173}$Yb atoms near an orbital Feshbach resonance in the presence of tunable population imbalances. Using realistic parameters, we identify the full phase diagram by independently tuning both ZEs of both channels, and find that with balanced population of closed channel, both the FFLO and Sarma superfluid phases are greatly enhanced in the BCS side by the strong inter-channel pair tunneling induced by spin-exchange interaction.
Importantly, because the robust BCS core in a harmonic trap center can be fully suppressed, not only the Sarma states with both one and two Fermi surfaces but also the FFLO state can be detected by measuring the polarization distributions of relevant band in current experiments.
Our results suggest another route to realize and detect the long-sought Sarma and FFLO states in alkaline-earth-like atomic gases.

\begin{acknowledgements} We thank Dr. Ren Zhang and Dr. Xibo Zhang for helpful discussions on experimental details. This work is supported by the National Key R$\&$D Program of China [Grants No. 2016YFA0301500 (W.L.) and No. 2018YFA0306501 (W.Z.)], the National Natural Science Foundation of China [Grants No. 11434015 (W.L.), No. 11434011 (W.Z.), No. 61835013 (W.L.), No. 11522436 (W.Z.), and No. 11774425 (W.Z.)], SPRPCAS (Grants No. XDB01020300 and No. XDB21030300), the China Postdoctoral Science Foundation [Grants No. 2018M631608 (D.Y)], the Beijing Natural Science Foundation (Grant No. Z180013), and the Research Funds of Renmin University of China (Grants No. 16XNLQ03 and No. 18XNLQ15).
\end{acknowledgements}

\appendix

\section{Grand thermodynamic potential and saddle-point equations}\label{append-a}

In grand canonical ensemble, the grand thermodynamic potential (in 3D uniform space) with respect to the normal state at zero temperature is calculated in standard formalism and explicitly divided into three parts: the thermodynamic potential associated with the open ($\delta G_o$) and closed ($\delta G_c$) channels ($j\equiv\{o,c\}$ hereafter),
\begin{align}\label{grandpot}
\delta G_j&=-\frac{M}{8\pi\hbar^2}\left(\frac{1}{a_{s+}}+\frac{1}{a_{s-}}\right)\Delta^2_j-
\sum_{{\bm k},\sigma}\xi_{{\bm k},j,\sigma}\Theta(-\xi_{{\bm k},j,\sigma})\nonumber\\
&+\sum_{\bm k}\left[ \frac{\Delta_j^2}{2\epsilon_{\bm k}}
+\bar\xi_{{\bm k},j}-E_{{\bm k},j}+\sum_{\pm}E_{{\bm k},j,\pm}\Theta\left(-E_{{\bm k},j,\pm}\right)\right],
\end{align}
and the interference energy between the two channels,
\begin{equation}\label{PairTunneling}
\delta G_{oc}=-\frac{M}{4\pi\hbar^2}\left(\frac{1}{a_{s-}}-\frac{1}{a_{s+}}\right)\Delta_o\Delta_c,
\end{equation}
which results in the inter-channel pair tunneling~\cite{ksun2013}, $\hbar$ is the Planck`s constant.
The saddle-point equations for the two superfluid (SF) order parameters and the magnitude of pair wavevector then read,
\begin{align}
\frac{4\pi\hbar^2}{M}\Delta_o R_o&=\left(\frac{1}{a_{s-}}+\frac{1}{a_{s+}}\right)\Delta_o + \left(\frac{1}{a_{s-}}-\frac{1}{a_{s+}}\right)\Delta_c,\nonumber\\
\frac{4\pi\hbar^2}{M}\Delta_c R_c&=\left(\frac{1}{a_{s-}}+\frac{1}{a_{s+}}\right)\Delta_c + \left(\frac{1}{a_{s-}}-\frac{1}{a_{s+}}\right)\Delta_o,\label{GapEqs}
\end{align}
and
\begin{equation}
|\bm{Q}|\Big( S_o+S_c \Big)=T_o+T_c,\label{SaddlepointEqu}
\end{equation}
where three specialized functions in each channel are defined as
\begin{align}
R_j&=\sum_{\bm{k}} \frac{1}{\epsilon_{\bm k}}-\frac{\Theta(+E_{\bm{k},j,-})-\Theta(-E_{\bm{k},j,+})}{E_{\bm{k},j}},\nonumber\\
S_j&=\sum_{\bm{k}} 1-\frac{\bar{\xi}_{\bm{k},j}}{E_{\bm{k},j}}\left( \Theta(+E_{\mathbf{k},j,-})-\Theta(-E_{\bm{k},j,+}) \right),\nonumber\\
T_j&=\sum_{\bm{k}} |\bm{k}|\cos(\theta_{\bm{k}})\left( \Theta(-E_{\mathbf{k},j,-})-\Theta(-E_{\mathbf{k},j,+}) \right).
\end{align}
In the expressions above, $\theta_{\bm k}$ is the angle between $\bm{Q}$ and $\bm{k}$, $\Theta(\cdot)$ is the Heaviside function. The energy dispersions of quasi-particles are $E_{\bm{k},j,\pm}=\pm (\hbar^2 \bm{Q} \cdot \bm{k}/M+h_j)+E_{\bm{k},j}$ with $E_{\bm{k},j}=\sqrt{\bar\xi^2_{\bm{k},j}+|\Delta_j|^2}$ and $\bar\xi_{\bm{k},j}=\epsilon_{\bm k}+\epsilon_{\bm Q}-\mu_j$.

Due to the intrinsic feature of the $^{173}$Yb orbital Feshbach resonance (OrbFR), its `{\it ground}' state must satisfy the following three constraints simultaneously:
1) It is an out-of-phase solution of the gap equations (Eqs.~\ref{GapEqs}) and the saddle-point equation (Eq.~\ref{SaddlepointEqu});
2) It gives the lowest energy $\delta G$;
3) The matrix of the second derivatives of $\delta G$ with respect to $\Delta_{o/c}$ and $|\bm Q|$ possesses only one non-positive eigenvalue.

\section{Quantum phase transition}\label{append-b}
\begin{figure}
  \centering
  \includegraphics[width=8.5cm]{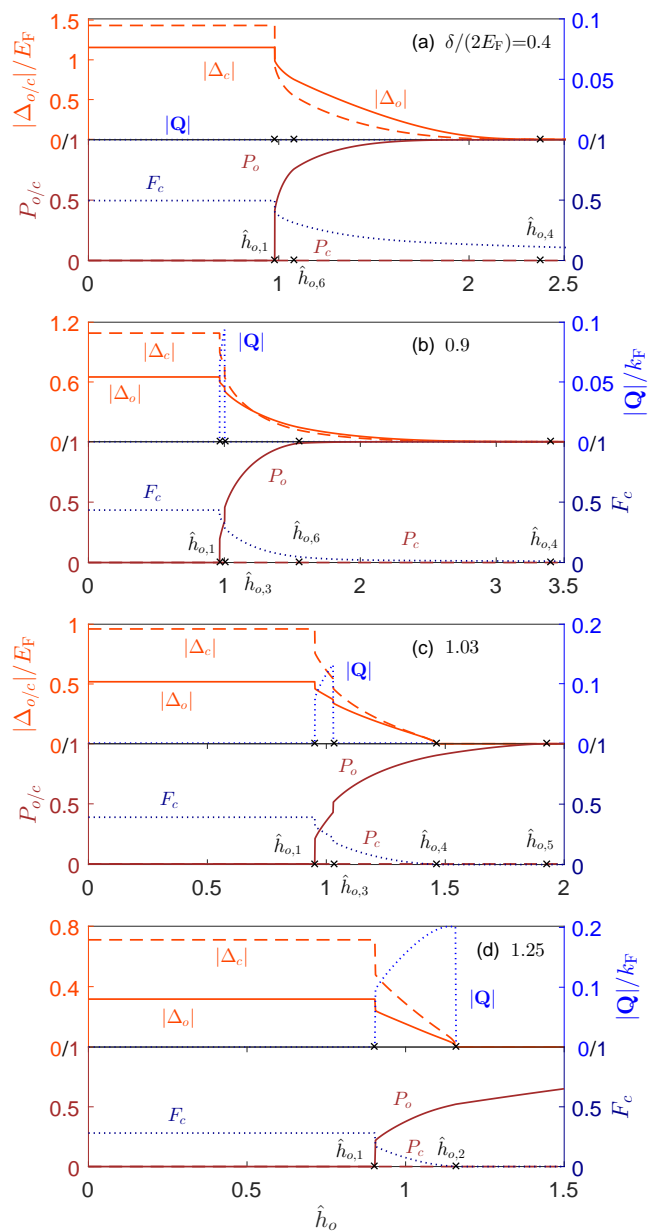}\\
  \caption{(Color online) Magnitudes of order parameters $\Delta_{o/c}$ (orange solid/dashed lines), wavevector of the FF state $|\bm{Q}|$ (blue dotted lines), polarization $P_{o/c}$ of each channel (brown solid/dashed lines), and the fraction of the closed channel $F_c$ (dark blue dotted lines) versus the dimensionless scaled ZE ${\hat h}_o$ as the energy detuning $\delta/(2E_{\rm F})$ equals to $0.4$ (a), $0.9$ (b), $1.03$ (c) and $1.25$ (d).
  The crosses in each panel label the critical points of phase transitions, other notations and settings are the same as in Fig.~\ref{Fig2}.}\label{Fig7}
\end{figure}

To further characterize various phases and quantum phase transitions discussed in Section~\ref{section-crossover}, we present the evolutions of several physical quantities with the dimensionless scaled ${\hat h}_{o}$ in Fig.~\ref{Fig7} for four typical values of $\delta$. In the BCS$_o$ phase with both $|\bm Q|=0$ and $P_o= 0$, we find that the closed channel is occupied macroscopically even in the BCS limit with $\delta \gg \delta_{\rm res}$ due to the inter-channel pair tunneling, as can be seen from the closed channel fraction $F_c$. In the two-band regime of $\delta/2\mu \lesssim 1$, the open and closed channels are approximately equally populated with $F_c \sim 0.5$, as shown in Figs.~\ref{Fig7}(a)-\ref{Fig7}(c). However, when $\delta/2\mu \gtrsim 1$, the system crosses over to the single-band model with $F_c$ significantly reduced from 0.5 as in Fig.~\ref{Fig7}(d). In the Sarma$_o$ phase with $|\bm Q|=0$ and $P_o\neq 0$, a phase separation in momentum space is clearly observed and verified by the density distribution (see Appendix~\ref{append-c}), which results in detectable signatures of this state as discussed in the main text and Appendix D. In the FF$_o$ state with both $|\bm Q| \neq 0$ and $P_o\neq 0$, the wavevector $|\bm Q|$ increases roughly linearly with $h_o$~\cite{sheehy2007}. From the behaviors of these quantities, we conclude that in the mean-field level, the phase transition from the BCS$_o$ or Sarma$_o$ to the FF$_o$ phase is first-order (thick dotted lines in Fig.~\ref{Fig2} )~\cite{strack2014,boettcher2015-1}, while the one from the Sarma$_o$ or the FF$_o$ to the normal state (FP$_o$ or PP$_o$) is second-order (thick solid lines in Fig.~\ref{Fig2}). It is remarkable that the polarization $P_c=(N_{c,\downarrow}-N_{c,\uparrow})/(N_{c,\downarrow}+N_{c,\uparrow})$ in the closed channel is always zero in all phases including the FF$_o$ state due to the large energy gap $|\Delta_c|$ and $h_c=0$.

\section{momentum distributions of Sarma SF}\label{append-c}
In alkaline-metal atomic gases, both the BCS and Sarma SF states are homogeneous in spatial space. However in contrast to the BCS state, the Sarma state is usually stabilized in the presence of large Zeeman energy (ZE) by reducing the magnitude of the SF order parameter $|\Delta|$ to become smaller than the corresponding ZE $| h |$. As a result, this state is fully polarized around the chemical potential and fully paired in the remained momentum space even at zero temperature. In other words, the Sarma state is a phase separation state in momentum space~\cite{vwliu2003,forbes2005}.
Specifically, if $|\Delta|<h<\sqrt{\mu^2+|\Delta|^2}$, a fully-polarized shell is formed in momentum space between the two Fermi surfaces (FSs) at $k_{{\rm F} \pm}=\sqrt{2M(\mu \pm \sqrt{h^2-|\Delta|^2})}/\hbar$ where quasi-particles are excited.
While if $h\geq\sqrt{\mu^2+|\Delta|^2}$, a fully-polarized ball can be seen below the FS with $k_{{\rm F} +}=\sqrt{2M(\mu + \sqrt{h^2-|\Delta|^2})}/\hbar$.

\begin{figure}
  \centering
  \includegraphics[width=8.5cm]{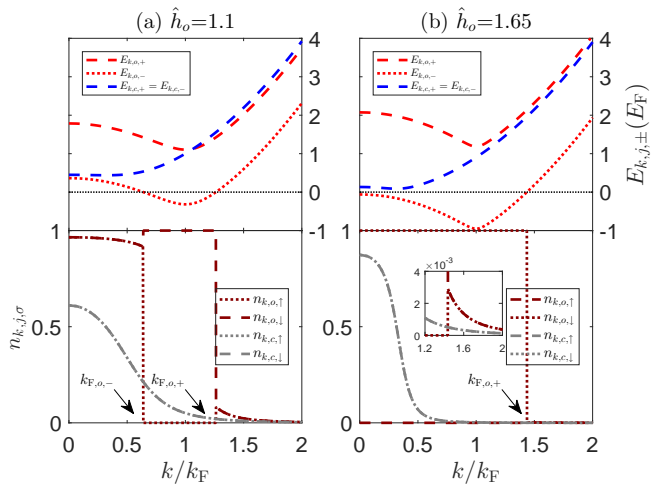}\\
    \caption{(Color online) Dispersions (upper panels) of quasi-particle excitations $E_{{\bm k},j,\pm}$ and density distributions (lower panels) of bare atoms $n_{{\bm k},j,\sigma}$ of the Sarma$_o$BCS$_c$ state in momentum space for (a) ${\hat h}_o=1.1$ and (b) $1.65$. The former (later) case corresponds to the Sarma state with two (one) FS(s). The inset in the lower-right panel is a zoom-in plot around the FS $k_{{\rm F},o,+}$, where $k_{{\rm F},o,\pm}$ is the Fermi wavevector. Here $\delta/(2E_{\rm F})=0.9$ and $k\equiv |\bm k|$. Other settings are the same as in Fig.~\ref{Fig2}.}\label{Fig8}
\end{figure}

In alkaline-earth-like $^{173}$Yb atomic gases, depending on the values of $h_o$ and/or $h_c$, the Sarma state can be stabilized in the open or the closed channel.
Here we take the Sarma$_o$BCS$_c$ state as an example with tunable $h_o$ and fixed $h_c=0$. In Fig.~\ref{Fig8} we show the typical dispersions of quasi-particle excitations $E_{{\bm k},j,\pm}$ and the corresponding momentum distributions of bare atoms $n_{{\bm k},j,\sigma}=\langle {\hat c}^\dagger_{{\bm k},j,\sigma} {\hat c}_{{\bm k},j,\sigma} \rangle$ in both channels. Here, $k_{{\rm F},o,+}=\sqrt{2M(\mu_o + \sqrt{h_o^2-\Delta_o^2})}/\hbar$ for $h_o\geq|\Delta_o|$, $k_{{\rm F},o,-}=\sqrt{2M(\mu_o - \sqrt{h_o^2-\Delta_o^2})}/\hbar$ for $|\Delta_o|<h_o<\sqrt{\mu_o^2+\Delta_o^2}$, and $k_{{\rm F},o,-}=0$ for $h_o\geq\sqrt{\mu_o^2+\Delta_o^2}$.

\section{Density distributions in a harmonic trap}\label{append-d}
In the presence of an external 3D isotropic harmonic trap, the characteristic length of the trap is much larger than other length scales, thus the local density approximation works well. By replacing the  chemical potential $\mu_j$ in Eq.~(\ref{H}) in Section~\ref{section-model} by the local chemical potential $\mu_j({\bm r})=\mu_j-V_{\rm ext}({\bm r})$, we get the local density,
\begin{align}
n_{j,\uparrow/\downarrow}({\bm k},r)=&\frac{1}{2} \bigg[ \left( 1+\frac{{\bar \xi}_{{\bm k},j}(r)}{E_{{\bm k},j}(r)}\right)\Theta(-E_{{\bm k},j,\pm}(r))\nonumber\\
&+\left( 1-\frac{{\bar \xi}_{{\bm k},j}(r)}{E_{{\bm k},j}(r)}\right)\Theta(+ E_{{\bm k},j,\mp}(r)) \bigg],
\end{align}
The global chemical potential $\mu$ is constrained by the total number of atoms $N$.

\begin{figure}
  \centering
  \includegraphics[width=8.5cm]{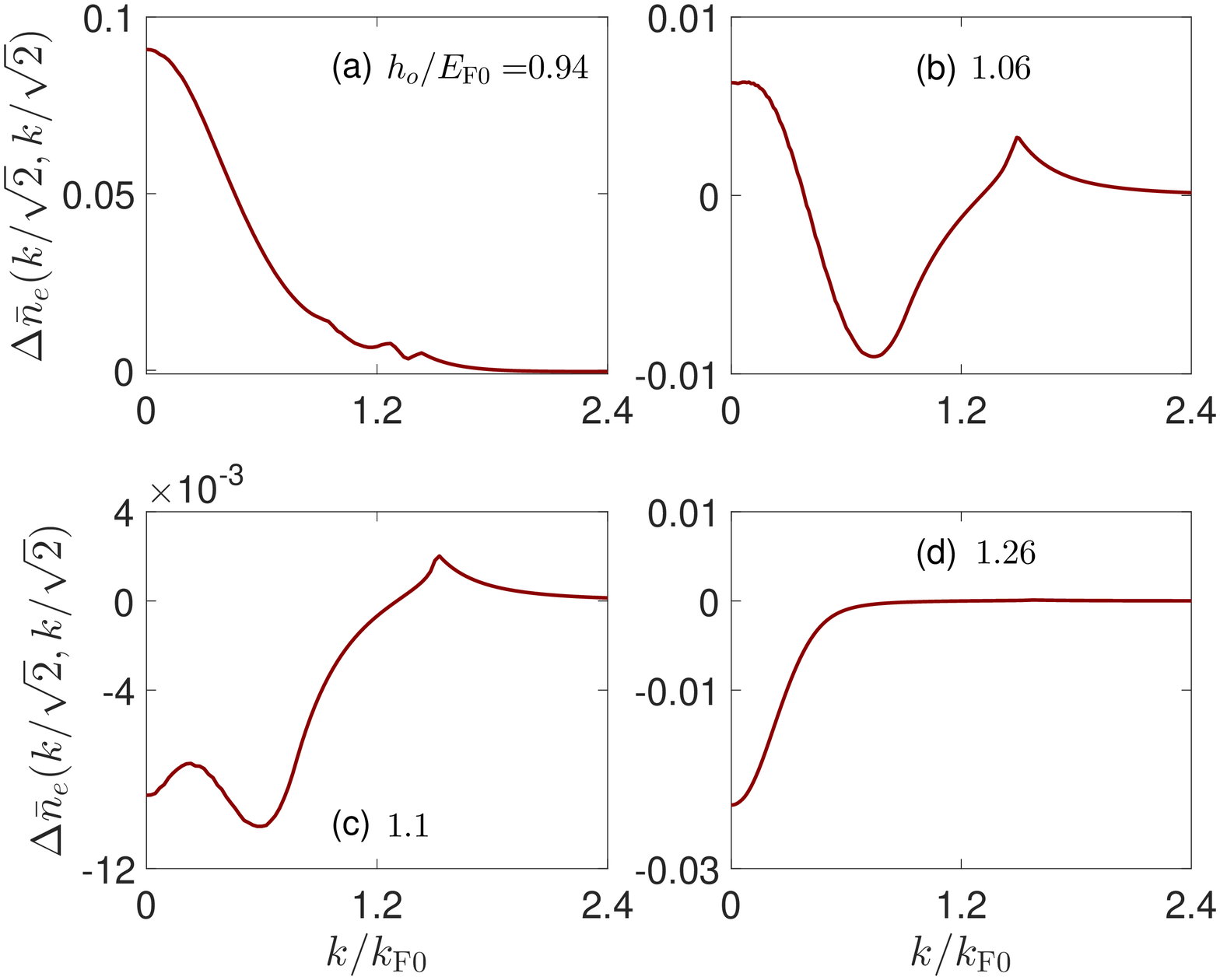}\\
  \caption{Polarization of $|e\rangle$ state $\Delta {\bar n}_e(k_x,k_y)$ when the FF$_o$ wavevector ${\bm Q}$ (if nonzero) is aligned along the \textit{z} axis, saying $\theta=0$. Other parameters are the same as in Fig.~\ref{Fig6} of main text.}\label{Fig9}
\end{figure}

Comparing with the Fig.~\ref{Fig6}(a-d), here we show the signatures of Sarma state in $\Delta {\bar n}_e(k_x,k_y)$ which can be obtained by spin-selective time-of-flight imaging technique. As we can see, the contrast of $\Delta {\bar n}_e(k_x,k_y)$ in Fig.~\ref{Fig9} becomes even larger and more visible
than the one of ${\bar n}_{e,\uparrow}(k_x,k_y)$ especially around $h_o\approx E_{{\rm F}0}$ (Fig.~\ref{Fig6}(b1,b2) and Fig.~\ref{Fig9}(b)) where the density distributions between $|o,\uparrow\rangle$ ($|e,\uparrow\rangle$) and $|c,\downarrow\rangle$ ($|e,\downarrow\rangle$) are comparable. Finally, the signature of Sarma state almost disappears in Fig.~\ref{Fig9}(d) where it is dominated by the density of closed channel.

\bibliography{mybib}

\end{document}